# Routing Technique Based on Clustering for Data Duplication Prevention in Wireless Sensor Network

Boseung Kim
Department of Computing
Soongsil University
Seoul, South Korea

Huibin Lim
Department of Computing
Soongsil University
Seoul, South Korea

Yongtae Shin
Department of Computing
Soongsil University
Seoul, South Korea

*Abstract*— Wireless Sensor Networks is important to node's energy consumption for long activity of sensor nodes because nodes that compose sensor network are small size, and battery capacity is limited. For energy consumption decrease of sensor nodes, sensor network's routing technique is divided by flat routing and hierarchical routing technique. Specially, hierarchical routing technique is energy-efficient routing protocol to pare down energy consumption of whole sensor nodes and to scatter energy consumption of sensor nodes by forming cluster and communicating with cluster head. but though hierarchical routing technique based on clustering is advantage more than flat routing technique, this is not used for reason that is not realistic. The reason that is not realistic is because hierarchical routing technique does not consider data transmission radius of sensor node in actually. so this paper propose realistic routing technique base on clustering.

*Keywords-Wireless Sensor Networks, Clustering*

I. INTRODUCTION

The recent technology of wireless communication and electronics makes it possible to develop multi-functional sensor-nodes of small sizes, which enable communicating between short distance, with such low costs, and relatively a little amount of electrical power.

Network Protocol is one of the technical factors which to organize the wireless network. As the wireless network has some factors to be overcome, which is not the case for the traditional networks, it is important to understand this traits in advance before designing wireless network.

Among these traits, It is the requirement for efficient utilization of the energy resources that should be regarded the most important for the reflection into network protocol. If network protocol operates in the surroundings of wireless sensor network where communications occur frequently without any consideration for the resources of energy, it can interfere with the operation of wireless sensor network by causing separation, isolation, interruption etc. of network[1,2,3].

Routing Protocol of wireless sensor network diverges largely into plane routing Protocol and the hierarchical routing protocol. Plane routing protocol regards the whole network as one region, enabling all the nodes to participate in; It is the technique to have the multi-hop routing as its trait. And, hierarchical routing protocol is the technique to grant the role of heads to the specific nodes in each region by dividing network into many regions based upon cluster[10].

This paper suggests RTBC(Routing Technique Based on Clustering), routing protocol based on cluster, organizing network as per a cluster, which can grasp the traits of communication happening in the surroundings of wireless sensor network, and control the resources of energy in terms of protocol. RTBC sets up a route between sink and cluster head by using the data values of sensor nodes distributed randomly, suggesting the technique for each member of nodes to transmit efficiently sensing information in cluster organized of cluster heads selected randomly like LEACH[4].

The structure of the paper is as follows: Section 2 in the paper discusses subjects to be considered of hierarchical protocol of sensor network, and analyzes various traits, weak and strong points. Section 3 suggests RTBC, routing protocol based on cluster, which can transmit efficiently sensing information by organizing network as per a cluster. Section 4 suggests the devices to realize the simulation of RTBC, and analyzes the efficiency of protocol suggested. At last, Section 5 summarizes the contents of paper and suggests the direction of research on the field later on.

II. RELATED STUDY

Flooding is the traditional technique being used in wireless sensor network. Flooding is the technique for them to repeatedly transmit the packet to their adjacent nodes in case that the nodes receiving packet are not the last, or can not reach the most numbers of hops of packet. However, it has three(3) problems of double message, double sensing, and efficiency of energy which should be overcome so that flooding can be used in wireless network.

SPIN(Sensor Protocols for Information via Negotiation)[5] is the protocol to transmit sensing information to many nodes via three(3) steps of negotiation in order to improve double message, double sensing, and efficiency of energy, which was pointed out as weak points of flooding. The message of SPIN includes meta-data which is the concise data on sensing information. It decides the double message and double sensing information via negotiation before carrying out the transmittance of message. This trait of meta-data can control network protocol, which is distinctive for SPIN.

DD(Direct Diffusion)[6] is the routing technique focusing data based upon question-broadcasting of Sink, which can transmit sensing information on the specific region to random nodes. DD transmits sensing information after setting up a route reversely from the targeted region to source nodes via three(3) steps.

LEACH(Low-Energy Adaptive Clustering Hierarchy)[4] is the routing protocol based on clustering for the purpose of dispersing the energy of nodes which organize network by themselves. In LEACH, selected cluster heads collect sensing information from member nodes of cluster, and transmit it Sink by itself.

It is pointed out as traits of this technique that LEACH makes cluster heads, which functions as energy intensive, circulate at random in order to distribute energy waste equally to all the sensor-nodes in network, and collect and manage data of cluster from cluster heads for saving the cost of communication. But it is difficult to be applied to the real situation considering that all the nodes, which are selected as cluster heads, should communicate directly with Sink.

## III. ROUTING TECHNIQUE BASED ON CLUSTERING

### A. RTBC

Even though the algorithm of hierarchical RTBC(Routing Technique Based on Clustering) has more strong points than the algorithm of plane routing, it is not used because it is unrealistic. In order to apply the algorithm of hierarchical routing to the real model, radius of transmitting data of senor-nodes should be taken into consideration. IEEE 802.15.4, known as the criteria of sensor network, defines radius of transmitting data of senor-nodes as 10m[7]. MICA2, which is being used most commonly as sensor-node, also rules the maximum radius as 10m[8]. This paper also limits the maximum radius of transmitting data of nodes to 10m.

This chapter suggests RTBC(Routing Technique Based on Clustering) using sensor-nodes which have the limiting radius of transmittance. Like LEACH, RTBC selects cluster heads in between nodes by the equal times based on probability, and organize cluster based upon the selected cluster head.

*1) Selecting Cluster Heads:* It is the first priority to obtain the information of sensor-nodes, which are distributed randomly at first in order to select cluster heads. So Sink transmits questioning message to sensor-nodes, which are one(1) hop away. Sink is able to count the numbers of sensor-nodes distributed at random as each nodes transmit its hop-count and ID to Sink in the responses to the questioning message. Using the sensor-nodes to be obtained this way, like LEACH[4], Sink selects cluster heads in between nodes by the equal times based on probability so that the energy waste between nodes in network.

*2) Organizing Cluster:* Like in LEACH, cluster heads are selected randomly by Sink nodes. The cluster heads, selected randomly, select nodes which reach 5% of the entire nodes. In the process, the node selected as cluster head is to be received the cluster head ID(CHID) from Sink. The selected cluster head should organize cluster by notifying the adjacent nodes that it is the cluster head via ADV message.

The node, which received *ADV* message, organizes cluster by modifying its node information, and transmitting *REP* message later on. The message for organizing cluster is as follows.

TABLE I. MESSAGE FOR ORGANIZING CLUSTER

| Message | Explanation |
|---|---|
| *ADV* Message data | |
| $CH_{ID}$ | Node' characteristic ID<br>Data distinctive from other cluster head |
| $SN_{ID}$ | ID of node to transmit *ADV* message<br>Data to response *ADV* message to node to which message was sent |
| $Hop_{CH}$ | Hop count of transmitter<br>Data for direction to cluster head |
| Sense node data | |
| $My\_CH_{ID}$ | ID of cluster head belonging to |
| $DN_{ID}$ | ID of node to which data is to be sent |
| $Hop_{CH}$ | Own hop count value |
| *REP* Message data | |
| $RN_{ID}$ | ID of Node to respond to *ADV* message to<br>Same with $DN_{ID}$ of each node data |
| $Hop_{CH}$ | Own hop count value |

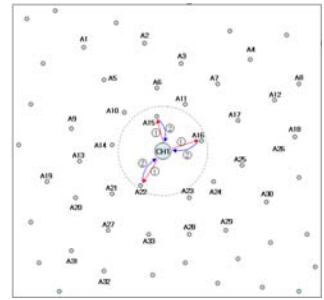

Figure 1. Cluster organizing and defined route within cluster-1

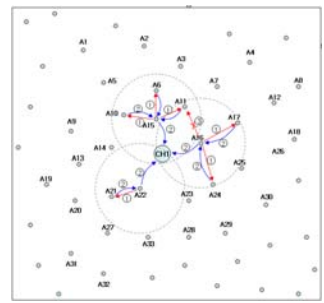

Figure 2. Cluster organizing and defined route within cluster-2

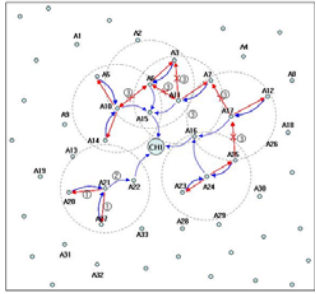

Figure 3. Cluster organizing and defined route within cluster-3

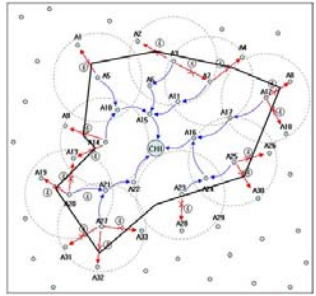

Figure 4. Cluster organizing and defined route within cluster-4

In [Figure 1], the selected cluster head *CH*1 transmits *ADV* message(*CH*1, *CH*1, 0) shown as ① to the nodes A15, A16, A22 which are one(1) hop away. The nodes which received the message, define $CH_{ID}$ of *ADV* message as the cluster head ID, *CH*1 which they belong to, and also define the sender node as *CH*1, their own $DN_{ID}$. And they set 1 for their $Hop_{cnt}$ value by adding 1 to the received $Hop_{cnt}$ value 0. Shown as ② in response to *ADV* message, each node transmits *REP* message(*CH*1, 1) which is received by the cluster head *CH*1, responding node of *REP* message. In this way, the sensor-nodes have direction to the cluster head.

[Figure 2] shows how the nodes A15, A16, A22 transmit again *ADV* message to the adjacent nodes. A15 re-transmits *ADV* message to the neighboring nodes A10, A6, A11. Then, shown as ① of [Figure 1], *ADV* message transmits the value of (*CH*1, *A*15, 1). And like in[Figure 1] A15, A16, A22 do, A10, A6, A11, which received this message, define their own information of sensor-nodes, and also, transmit *REP* message(*A*15, 2) to A15. Then, A15 decides whether the hop count value($Hop_{cnt}$) is 0. If hop count value($Hop_{cnt}$) is not 0, which means that this node plays a role of the mid-node, A15 re-transmits *REP* message(*CH*1, 1) shown as ②. At last, the cluster head is able to recognize the node within its cluster by receiving this value.

[Figure 3] shows node A15, [Figure 2] shows the third stage to transmit *ADV* message. ③ shows the competition between sensor-nodes, or non transmittance of message owing to the nodes receiving *ADV* message. In case of A10, *ADV* message can be transmitted to A5, A14, but in case of A6, it can have the same hop-count value with A6 through *ADV* message transmitted from A15. Therefore, *ADV* message transmittance does not happen mutually because they are judged as the same level of nodes. Even though A6 and A11 transmit simultaneously *ADV* message, the one that arrives first can bring A3 to its route. As shown in the picture above, A6 preoccupies A3 as it is closer than A11.

Sensor-nodes, which extended *ADV* message, meet the sensor-nodes with different *CH*1. In ④ of [Figure 4], A31, A32, A33 do not receive *ADV* message which A27 transmitted as A31, A32, A33 have different cluster head from A27. This part form the boundary in between clusters, organizing cluster of *CH*1 as shown in the picture above.

*3) Routing Within Cluster:* After each member node organized cluster, cluster head *CH*1 defined the imaginary routs of nodes for the direction of itself based upon the each belonging node data. Though node *CH*1 recognizes only nodes, which are one(1) hop away, each node also is connected together by this low level of information. Therefore, the node, which has an event, can transmit data by *CH*1 following the imaginary route set up shown as above. In case above, cluster head does not need to define route by transmitting a questioning message to the node with the event. Also to define the only route is because cluster heads change regularly; it is more efficient to maintain the transmittance of data within cluster via the defined route rather than to re-define the route according to events.

*4) Routing out of Cluster:* Each sensing information, which was received by cluster head from all the nodes, makes double data of cluster head as one, and checks the condition of each node by transmitting multi-hop and transmits it to Sink later on. Sink node should transmit regularly the interest message to network for the sake of communication between cluster head and Sink. Then, interest message is to be transmitted to the whole network from Sink node, and each node existing in the network recognizes the energy and numbers of hops of neighboring nodes by using this message. When transmitting data to Sink node, cluster head defines the nodes, of which condition of energy is good, and the number of hops is small as the receiving nodes among its neighboring nodes' tables, and transmits data later on. Also, the nodes receiving the data of cluster head transmit data in the same manner. Considering that it is difficult for cluster head to communicate directly with Sink node, the routing technique of cluster head suggested in this paper uses routing based upon the neighboring energy and numbers of hops directing to Sink. The suggested routing technique does not maintain a special routing-route for routing, but it is easy to use as is routing to the neighboring nodes having minimum of hops to Sink node. Additionally, cluster head can use the shortest distance to transmit data to Sink node.

## IV. REALIZATION AND ANALYSIS ON EFFICIENCY

### A. Evaluation Model for RTBC Efficiency

For evaluating efficiency, routing technique based upon clustering with limited radius of transmittance was realized by C++, and the related factors are decided to define the related environment. For simulation make-up, assuming N units of sensor-nodes are to form in space of a regular square coordinates, movability and additional nodes was not considered. Also, each node has the same trait, and begins from the same condition. The nodes selected as each cluster head is the same nodes as well.

In the process of experiment, it was noticeable that it can screen double data through clustering. To measure the efficiency of utilizing energy, it was done to compare the average amount of energy waste of the entire network according to the event node, to the one of established plane routing by changing the cycle of organizing cluster. Accordingly the amount of energy waste was measured.

*1) Definition of Environment Factor:* As shown in Table 2, the size of network was limited to 100m x 100m. The numbers of sensor-nodes are to be used for recognizing the numbers of nodes with no errors which the simulation has in the size of network. Each node, which has event, was occurred at random from 100 units to 500 units. The range of sensing was defined as 10m based on the distance of nodes having limiting radius, and the maximum distance between each node was limited to 5m.

Assuming that the coordinates defines as (50, 0), the energy of each node distributed at first, transmitting and receiving energy was also defined as in Table 2.

TABLE II. SURROUNDING FACTOR FOR SIMULATION

| Surrounding factor | | Value of setting up | |
|---|---|---|---|
| Size of network | | *100m x 100m* | |
| Sensor-node | Unit of nodes | *50,100,150,200,250,300* | |
| | Unit of event nodes | *100,200,300,400,500* | |
| | Unit of round | *50,100,200* | |
| | Range of sensing | *10m(=1hop)* | |
| | Minimum distance between nodes | *5m* | |
| Sink | Position | *(50,0)* | |
| Energy | Value at first | *100unit* | |
| | Transmitting | *1unit (data)* | *0.25 (interest)* |
| | Receiving | *1unit (data)* | *0.25 (interest)* |

## B. Evaluation and Analysis on RTBC

In the experiment, the numbers of nodes of each cluster head in the network which does not have the isolation of nodes, and having 300 units of nodes, were compared. In average, the cluster forms the stable shape with 20 units of nodes.

In the simulation, DD and RTBC were compared respectively 10 times under the same condition in sensor network having 300 units of nodes. Also, the frequency for organizing cluster was experimented by changing the occurrence of the event nodes as 50, 100, 200. This shows which frequency of occurrence of the event nodes is the most efficient; This is because organizing cluster needs more energy waste than non hierarchical techniques.

The numbers of each message and the consuming amount of total energy were compared. The horizontal axis stands for the units of event, and the vertical axis of coordinators stands for the numbers of interest message according to the occurrence. Looking over RTBC(50), RTBC(100), RTBC(200) which define the frequency to organize clustering respectively as 50, 100, 200, the demanding interest message is higher than DD which is non hierarchical routing technique when RTBC is set to 50 at the lowest. But when RTBC is set over 100, the interest message is lower than DD, or at the almost same level.

When comparing the numbers of message presented by the simulation, the message technique using clustering is shown as effective to protect double data. Also its effectiveness on the whole is as follows.

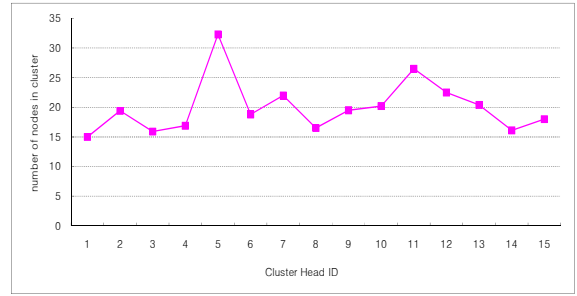

Figure 5. Number of nodes within cluster

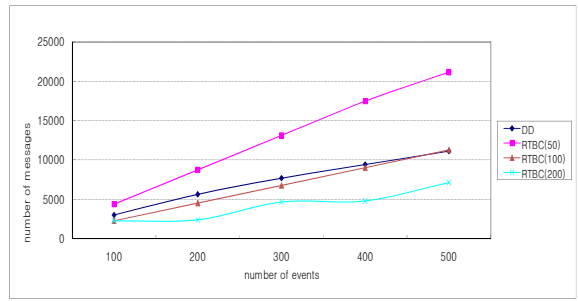

Figure 6. Comparison of number of interest messages

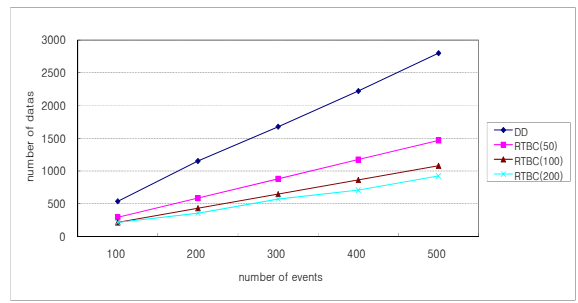

Figure 7. Comparison of number of datas

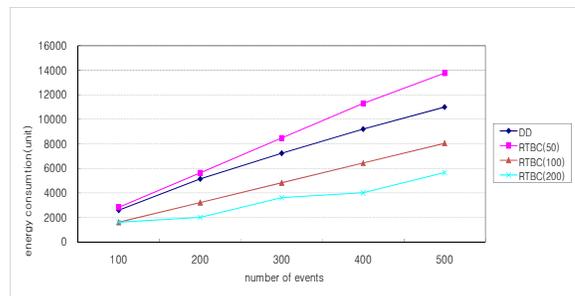

Figure 8. Comparison of the total amount of energy consumption

*C. Result of Evaluating Efficiency of RTBC*

RTBC, the hierarchical routing technique for the wireless sensor network was defined to 10m, which is the range of receiving and transmitting, and the realistic and practical technique was suggested through routing within and out of cluster. The result of the experiment above can induce several consequences.

First, in case of using the hierarchical technique applied with clustering in the sensor network, it was possible to save entirely the energy waste as well as to consume the energy efficiently through the equal distribution of energy.

Second, the numbers of interest message to the nodes with the occurrence of event also decreased. This can help to improve the efficiency of energy by over 18% on the average through the experiment. By transmitting interest message received from Sink, the cluster head can also prevent double message.

Third, it can not only improve the credibility of transmitting data, but help to save the energy waste nationwide to prevent double data of cluster. RTBC was proven as the efficient routing technique by preventing about 58% of double message transmittance.

Fourth, it is possible to organize the realistic clustering by using the sensor-nodes having the limiting range, which means that it is possible to use the trait of sensor-nodes based upon the communication of low electrical power. But it needs to define properly the frequency to organize each cluster so that this can be possible.

## V. CONCLUSION

In wireless sensor network, it is more important to preserve the energy of nodes for organizing the continuous network than to consider efficiency owing to the trait of applications program and limitation of hardware. Also, collecting the sensing information should be easy. These traits can be applied to the network protocol, and the protocols of Flooding, SPIN, DD, and LEACH were suggested by the former research. But even though LEACH using the hierarchical routing can have a lot of strong points by sensing double data or managing regionally data transmittance, it is not efficient because it is not appropriate for the sensor-nodes having the limiting range.

Therefore, the network protocol, which realistically has limiting range of transmittance, and can sense double data and manage it regionally compared with non hierarchical routing, is required. So this paper suggests RTBC, routing technique based on clustering for preventing double data, which recognizes diachronic trait in the surrounding of wireless sensor network; For this purpose, comparison between RTBC and the established non hierarchical routing technique was done by defining the process of organizing cluster, routing within cluster, and routing out of cluster.

Through the simulation, it was experimented preventing double data of RTBC and analyzing the efficiency of managing data regionally, also it was induced that the realistic routing technique based on clustering is possible, and superior from comparison and evaluation with DD.

Based on the result of this research, it is well expected that the realistic routing technique will be able to used widely through preventing double data through clustering and managing data regionally.

AUTHORS PROFILE

B. Kim. Author is with the Department of Computing, Ph.D. course, Soongsil University, Seoul, Korea. His current research interests focus on the communications in wireless sensor networks (e-mail:bskim@cherry.ssu.ac.kr).

H. Lim. Author is with the Department of Computing, M.Sc. course, Soongsil University, Seoul, Korea. His current research interests focus on the communications in wireless sensor networks (e-mail: jhlee@cherry.ssu.ac.kr).

Y. Shin. Author was with the Computer Science Department M.Sc. and Ph.D., University of Iowa. He is now with the Professor, Department of Computing, Soongsil University. (e-mail: shin@ssu.ac.kr).